\documentstyle[12pt,epsf]{article}

\newcommand{\bra}[1]{\langle  {#1}  \vert }
\newcommand{\ket}[1]{\vert {#1} \rangle }

\begin{document}
\input epsf

\begin{flushright}
SLAC--PUB--8267\\
September 23, 1999
\end{flushright}
\bigskip\bigskip

\thispagestyle{empty}
\flushbottom

\centerline{{\Large\bf 
Quarks, Gluons and Frustrated Antiferromagnets}
\footnote{Work supported in part by Department of Energy contracts
DE--AC03--76SF00515 and DE--AC02--76ER03069}}
\vspace{22pt}
\centerline{\bf Marvin Weinstein}
\vspace{8pt}
\centerline{\it Stanford Linear Accelerator Center}
\centerline{\it Stanford University, Stanford, California 94309}
\vfill
\begin{center}
ABSTRACT
\end{center}

The Contractor Renormalization Group method (CORE) is used to establish
the equivalence of various Hamiltonian free fermion theories and a class of
generalized frustrated antiferromagnets.  In particular, after a detailed
discussion of a simple example, it is argued that a generalized frustrated
$SU(3)$ antiferromagnet whose single-site states have the quantum
numbers of mesons and baryons is equivalent to a theory of free massless quarks.
Furthermore, it is argued that for slight modification of the couplings
which define the frustrated antiferromagnet Hamiltonian, the theory becomes
a theory of quarks interacting with color gauge-fields.

\vfill

\begin{center}
Submitted to Physical Review D.
\end{center}
\vfill  
\newpage

\section{Introduction}
\label{intro}

It may seem surprising that a Hamiltonian lattice theory whose
single-site states only have the quantum numbers
of mesons and baryons can be equivalent to a theory of free massless quarks,
but it is true.  I will show that this result
follows directly from the application of the Contractor Renormalization
Group (CORE) method\cite{COREpaper} to an appropriately chosen free fermion theory.

The original motivation for the computation I will present was the desire to
apply CORE to lattice quantum chromodynamics (QCD) and show that
the picture which emerged from older strong-coupling calculations\cite{chiralsymmbb}
also applies to the weak coupling regime.  The new feature of CORE which allows
this question to be dealt with nonperturbatively is that CORE, as opposed
to earlier Hamiltonian real-space renormalization group approaches, allows one
to retain only gauge-invariant block-states (i.e., states in which no flux
leaves a block) and still define a non-trivial renormalization group transformation.  

Fluxless states are of particular importance to the strong coupling
limit of a lattice gauge-theory because every link which carries nonvanishing flux
contributes an energy proportional to $g^2$, so that fluxless states have the lowest
energy.  Moreover, if a lattice theory allows for single-site color-singlet states
(e.g., theories which exhibit explicit fermion doubling, Wilson fermions, domain-wall fermions,
or theories based upon variants of the SLAC-type of derivative) the number of degenerate
fluxless states is huge and in the limit $g^2 \rightarrow \infty$ these states will all 
have zero energy.  In the case of lattice QCD,
single-site meson states (i.e., color-singlet quark anti-quark states), single-site
baryon states, and single-site multi-meson and baryon states consistent with the
exclusion principle, are all of this type.
The huge degeneracy among these fluxless states is lifted in order $1/g^2$ and,
for a nearest-neighbor derivative, perturbing in $1/g^2$ leads to an
effective Hamiltonian which has the form of a generalized Heisenberg antiferromagnet.
An immediate consequence
of this result is that chiral $SU(3)\times SU(3)$ is spontaneously broken in the
groundstate of this theory; another general result
is that the theory has an approximate $SU(12)$ symmetry which is broken if one
adds next-to-nearest neighbor terms to the fermion derivative.

Although it is attractive to rewrite strong-coupling QCD in terms
of states which have the quantum numbers of mesons and baryons, establishing the
relevance of the chiral-symmetry prediction and the approximate $SU(12)$ results to the
continuum, small $g^2$, limit is problematic.  CORE allows one
to systematically study this question by truncating
the Hilbert space to this set of strong-coupling 
states, obtaining a renormalized Hamiltonian which is valid for all values of $g$.
The important question which must be answered is whether 
truncation to this set of states biases the computation and incorrectly
forces the strong coupling results of confinement and chiral symmetry breaking.
One way to show that this is not the case is to apply the same truncation scheme
to free fermion theory and show that it leads to a {\it renormalized Hamiltonian\/} which has the
same physics as the global color-singlet sector of the free theory.
While the general theorem on which CORE is based guarantees this result will hold if
the retained states have a non-trivial overlap with the
relevant low-lying states of the free-field theory, a-priori nothing forces
this overlap to be nonvanishing; establishing this fact
requires a calculation.

This paper does the requisite calculation for a theory of  
free massless fermions in $1+1$ space-time dimensions with a nearest
neighbor fermion derivative.  It will be clear that the extension of this calculation to
higher dimensions and other derivatives is straightforward.
While the nearest-neighbor theory exhibits species doubling
and is anomaly free, a fact which makes it useless for studying the physics
of theories such as the Schwinger model, it is simple to use and is relevant to the
question of whether or not truncating to the natural strong-coupling 
states makes it impossible to obtain the correct weak-coupling physics.
The explicit calculation shows that things work as expected for the
nearest-neighbor theory and that the {\it renormalized Hamiltonian\/} takes the
form of a {\it generalized \/} frustrated antiferromagnet
which, perforce, has all the physics of the charge zero sector of the
original free fermion theory.

Since there is no substantive difference between
the physics of free relativistic lattice fermions and the physics
of this unusual frustrated antiferromagnetic system (which 
doesn't, at the short distance level, have any relativistic fermions)
it is interesting to ask which theory is fundamental ?
Clearly, at this level there is no way to decide the
issue.  It will be apparent from the calculation that the same mapping
can be carried out for different choices of fermion derivative with similar
results.  In other words, the couplings will vary in strength but the general form
of the renormalized Hamiltonian will be the same.  From
a renormalization group point of view this says that there is a surface
in coupling constant space of the generalized frustrated antiferromagnetic system
where all Hamiltonians flow to the same free massless fermion fixed point.  
Furthermore, it will be clear that turning on gauge
fields in the original problem produces the same sort of Hamiltonian
with different coefficients. Thus, the gauge
theory is also hidden inside this same system or, in other words, all of these
different theories are different phases of the same generic 
Hamiltonian.  As a class, frustrated antiferromagnets are 
systems which have recently come to be of some interest because of their possible
connection to high-$T_c$ superconductors,\cite{OtherPapers}, so this identification
of a more general class of HAF's (for a specific
couplings) to a theory of free relativistic fermions,
or relativistic fermions interacting through a gauge-field, has interest 
beyond its application to QCD.

\section{A Brief Review of CORE}

The CORE method consists of two parts, a theorem which defines
the Hamiltonian analog of Wilson's exact renormalization group
transformation and a set of approximation procedures which render
nonperturbative calculation of the {\it renormalized Hamiltonian\/}
doable.

CORE replaces the Lagrangian notion of integrating out 
degrees of freedom by that of throwing away Hilbert space states.  This is
accomplished by defining a projection operator,
$P$, which acts on the original Hilbert space, ${\cal H}$ and whose
image is defined to be the space of {\it retained states\/}  ${\cal H}_{\rm ret} = P
{\cal H}$.   The foundation of the CORE approach is a 
formula which relates the original Hamiltonian, $H$, to the {\it renormalized
Hamiltonian\/} which has, in a sense which was made precise in
Ref.\protect\cite{COREpaper},  exactly the same low energy physics as
$H$.  This equation is
 \begin{equation}
        H^{\rm ren} = \lim_{t\rightarrow \infty}
            [[T(t)^2]]^{-1/2}\, [[T(t) H T(t)]] \, [[T(t)^2]]^{-1/2}   ,
\label{basicform}
\end{equation}
where $T(t) = e^{-t H}$ and where $[[O]]= P O P$ for
any operator $O$ which acts on ${\cal H}$. A similar formula can be written
to define the {\it renormalized\/} version of any other extensive operator.
The new, renormalized, operator is guaranteed to have the same matrix elements
between eigenstates of $H^{\rm ren}$ that the original operator had between
eigenstates of $H$.

\subsection{The cluster expansion}
\label{cluster}

Generally one cannot evaluate Eq.\ref{basicform}
exactly, however it is possible to nonperturbatively approximate the
infinite lattice version of $H^{\rm ren}$ to any desired degree of
accuracy.  This works because $H^{\rm ren}$, as defined in
Eq.\ref{basicform}, is an extensive operator and has the general form
 \begin{equation}
        H^{\rm ren} = \sum_{r=1}^{\infty} h^{\rm conn}(j,r) 
\label{hcluster}
\end{equation}
where each term, $h^{\rm conn}(j,r)$, stands for a set of {\it range-$r$
connected\/} operators based at site $j$, all of which can be
evaluated to high accuracy using finite size lattices.  The explicit definition of
the connected range-$r$ operator, $h^{\rm conn}(j,r)$, depends upon
the details of the truncation procedure.  In what follows I
will limit discussion to the case of a one-dimensional lattice, since this
is what I will need to discuss the free-fermion theory. A detailed discussion
of the general methodology can be found in Ref.\protect\cite{COREpaper}.
 
\subsection{The Approximation Procedure}
\label{approximations}

Three steps define the nonperturbative approximation scheme for
computing $H^{\rm ren}$:  first the truncation procedure;  second the
subtraction procedure used to convert the evaluation of
Eq.\ref{basicform} on finite sublattices to
the operators $h^{\rm conn}(p,r)$; third the method for evaluating the
$t \rightarrow \infty$ limit in Eq.\ref{basicform}, without explicitly
computing either $[[T^2(t)]]^{-1/2}$ or $[[T(t) H T(t) ]]$.   I will
heuristically review each of these steps in turn.

First some notation.  In what follows I deal with one-dimensional
spatial lattices whose sites are labelled by $-\infty \le j \le \infty$.
I assume that there are $N$-states corresponding to each site $j$ of the
lattice which I denote by $\ket{ \phi_\alpha(j) }_j$, where
$\alpha = 1\ldots N$.   A basis for the full Hilbert space ${\cal H}$
will be generated by taking tensor products of these $N$-states per site
over all sites $j$.  

CORE allows a wide choice of truncation procedures, however I
will limit myself to one which appears to work well in a
large number of cases. First, divide the lattice into disjoint
blocks $B_p$ each having $n_B$ sites and keep a small number states per block.
The way to choose which states to keep is to diagonalize the block-Hamiltonian
(i.e., that Hamiltonian obtained by restricting $H$ to only those terms
which are contained within any one of the blocks $B_p$) and throw away all but its $M$
lowest lying eigenstates, where $M < N^{n_B}$.  If we let $H_{B_p}$ denote
the block Hamiltonian and $\ket{\Psi_\alpha(p)}$ for $\alpha=1..M$
its eigenstates, then the projection operator $P$ is 
 \begin{eqnarray}
	P &=& \prod_p P_p \nonumber \nonumber \\
	P_p &=& \sum_{\alpha=1}^{M < N^{n_B}}  \ket{\Psi_\alpha(p)}\bra{\Psi_\alpha(p)}
\end{eqnarray}
Given $P$ it only remains to compute $H^{\rm ren}$.

Generally the lattice on which the renormalized Hamiltonian is defined is {\it
thinner\/} than the original lattice in that each site $p$ on the new
lattice corresponds to a block of sites $B_p$ on the old lattice, however
this need not be the case.  In the free-fermion case, to be
discussed in the next section, we will thin the states associated with
a single site of the original lattice and map the original theory into
an equivalent theory with the same number of sites but with a Hamiltonian
which has a very different form. 

To define the cluster expansion
begin by defining the range-1 term in $H^{\rm ren}$, $h^{\rm
conn}(p,1)$, to be
  \begin{eqnarray}
	h^{\rm conn}(p,1) &=& P\, H_{B_p}\,P \nonumber\\
	&=& \left( \sum_{\alpha=1}^M  E_\alpha\,\ket{\Psi_\alpha(p)}\bra{\Psi_\alpha(p)} \right)\, 
	 P^{\perp}_p 
\label{honeconn}
\end{eqnarray}
where $H_{B_p}\ket{\Psi_ \alpha(p)} =
E_\alpha \ket{\Psi_\alpha(p)}$ and where
\begin{eqnarray}
	P &=& P_p \, P^{\perp}_p  \nonumber\\
	P^{\perp}_p &=& \prod_{l \neq p} P_l  . 
\label{pperp}
\end{eqnarray}  
The range-2 connected operator $h^{\rm conn}(p,2)$  is defined by
subtracting $h^{\rm conn}(p,1)$ and $h^{\rm conn}(p+1,1)$ from
$H^{(2)}(B_p,B_{p+1})$, the operator obtained by evaluating the $t
\rightarrow \infty$ of Eq.\ref{basicform} for $H$ restricted to the two
adjacent blocks $\{ B_p, B_{p+1} \}$.  Note, in this case the notation
$[[O]]$ stands for $ P=P_p P_{p+1} O P_p P_{p+1}$.
The explicit definition of $h^{\rm conn}(p,2)$ is
 \begin{equation}
	h^{\rm conn}(p,2) =  P^{\perp}_{p,p+1}\,H^{(2)}(B_p,B_{p+1})\,P^{\perp}_{p,p+1}
	- h^{\rm conn}(p,1) - h^{\rm conn}(p+1,1)  .
\label{rangetwoconn} 
\end{equation}
where, in analogy to Eq.\ref{pperp} I define
 \begin{equation}
	P^{\perp}_{p,p+1} = \prod_{l \neq p,p+1} P_l
\end{equation}
Similarly, the range three operator $h^{\rm conn}(p,3)$ would be obtained
from the following formula
 \begin{eqnarray}
	h^{\rm conn}(p,3) &=& P^{\perp}_{p,p+1,p+2}\, H^{(3)}(B_p,B_{p+1},B_{p+2})\, 
	P^{\perp}_{p,p+1,p+2} - h^{\rm conn}(p,1)  -  h^{\rm conn}(p+1,1)  \nonumber \\
 & &\qquad  - h^{\rm conn}(p+2,1) - h^{\rm conn}(p,2)  - h^{\rm conn}(p+1,2)  
\label{rangethreeconn}
\end{eqnarray}
In this case one must subtract
the three different ways of embedding the connected range-1
computation and the two-different ways of embedding the connected
range-2 computation in the three block problem.
(As before,  $ H^{(3)}(B_p,B_{p+1},B_{p+2}) $ is the operator which
results when one restricts $H$ to the three adjacent blocks $\{B_p,
B_{p+1},B_{p+2}\}$ and then evaluates Eq.\ref{basicform}.)

Although it is possible to numerically evaluate Eq.\ref{basicform} for
any multi-block sublattice and extract the limit $t \rightarrow \infty$
by taking large vales of $t$ (see for example Ref.\protect\cite{COREpaper})
this is not necessary   The fact is that each term in the cluster
expansion can be computed from a knowledge of the tensor product states
(which span the space of {\it retained states\/} for the multi-block
problem and the eigenvalues and eigenstates of the corresponding
multi-block Hamiltonian.  A general proof of this assertion appears
in Ref.\protect\cite{COREpaper}, however
the basic ideas are summarized in the following theorem.

\bigskip

{\narrower
\noindent{\bf Theorem:}  Let $H_{\rm B}$ be a single block Hamiltonian
and let $P$ be the projection operator which corresponds to keeping
its $M$ lowest lying eigenstates $\ket{\Psi_\alpha}$;
furthermore, let $H$ denote the Hamiltonian of an $r$-block sublattice
and let the $M^r$ tensor product states formed from the states
$\ket{\Psi_\alpha}$ span the space of retained states. Then the
$t\rightarrow \infty$ limit of the equation which defines the
renormalized multi-block Hamiltonian can always be written as
 \begin{eqnarray}
 \label{hamtinf}
        H^{\rm ren} &=& \lim_{t\rightarrow \infty}
            [[T(t)^2]]^{-1/2}\, [[T(t) H T(t)]] \, [[T(t)^2]]^{-1/2} \nonumber \\ 
            &=&  R \, H_{\rm diag} \, R^{\dag} 
\end{eqnarray}
where $R$ is an $M^r\times M^r$ orthogonal matrix, $R^{\dag}$ its inverse
and $H_{\rm diag}$ is a diagonal matrix whose entries are the 
eigenvalues of those $M^r$ lowest lying eigenstates of $H$ which appear in the
expansion of the retained states. 

}
\bigskip

To clarify what is meant by the $M^r$ lowest lying eigenstates
which appear in the expansion of the {\it retained states\/}
I will consider two simple examples. The first example corresponds to the
simplest truncation procedure one can imagine, i.e. choosing $M=1$ and
truncating to a single state. In this case the theorem is trivial since
$M^r=1$ and so the space of retained states for that multi-block system
is one dimensional.  The fact that $R$ and $R^{\dag}$ are orthogonal
matrices means their single matrix element must be $1$ and so, as long
as the single retained state has an overlap with the ground-state,
$H_{\rm diag}$ must simply be the groundstate energy of the 
multi-block Hamiltonian. To prove this assertion it suffices
to rewrite Eq.\ref{basicform} as
 \begin{eqnarray}
	\ket{\Psi(t)} &=& {e^{-tH} \ket{\Psi} \over \sqrt{\bra{\Psi} e^{-2tH}
	\ket{\Psi}}} \nonumber \\
	H^{\rm ren} &=& \lim_{t \rightarrow \infty} \bra{\Psi(t)} H
	\ket{\Psi(t)} 
\end{eqnarray}
and then expand $\ket{\Psi(t)}$ in a complete set of eigenstates of $H$.

The second example, $M=2$ and $r=2$, exhibits all essential features of the general
case.  Obviously, in this case $M^r=4$ and so $[[T^2(t)]]^{-1/2}$ and $[[ T(t) H T(t) ]]$
are $4\times 4$ matrices, each of which becomes singular in the limit
$t\rightarrow \infty$, although their product is well defined.  To
understand why the product is well defined and has the form shown in
Eq.\ref{hamtinf} it is convenient to expand the four {\it retained states\/}
in terms of exact eigenstates of the two-block problem and on the basis of
this expansion, construct an orthogonal transformation $R$ which renders
the evaluation of the limit in Eq.\ref{hamtinf} straightforward.
If we denote the four retained states
as $\ket{\Psi_1}$, $\ket{\Psi_2}$, $\ket{\Psi_3}$  and $\ket{\Psi_4}$
we can write their expansion in terms of block eigenstates as
\begin{eqnarray}
\ket{\Psi_1} &=& a_0 \ket{\phi_0} + a_1 \ket{\phi_1} + a_2 \ket{\phi_2} + \ldots \nonumber\\
\ket{\Psi_2} &=& b_0 \ket{\phi_0} + b_1 \ket{\phi_1} + b_2 \ket{\phi_2} + \nonumber\ldots\\
\ket{\Psi_3} &=& c_0 \ket{\phi_0} + c_1 \ket{\phi_1} + c_2 \ket{\phi_2} + \nonumber\ldots\\
\ket{\Psi_4} &=& d_0 \ket{\phi_0} + d_1 \ket{\phi_1} + d_2 \ket{\phi_2} + \ldots
\label{twostateexp} 
\end{eqnarray}
where the states $\ket{\phi_n}$ correspond to eigenstates of the block
Hamiltonian with energies $\epsilon_n$.  Assume that the states are arranged in
the order of increasing energy, so that $\phi_0$ is the groundstate of the
block Hamiltonian, $\ket{\phi_1}$ the first excited state, etc.  

Applying $T(t)$ to each of these states we obtain
\begin{eqnarray}
T(t)\,\ket{\Psi_1} &=& a_0 e^{-t\epsilon_0} \ket{\phi_0} + a_1 e^{-t\epsilon_1} \ket{\phi_1}
+ a_2 e^{-t\epsilon_2} \ket{\phi_2} + \ldots \nonumber \\
T(t)\,\ket{\Psi_2} &=& b_0 e^{-t\epsilon_0} \ket{\phi_0} + e^{-t\epsilon_1} b_1 \ket{\phi_1}
+ b_2 e^{-t\epsilon_2} \ket{\phi_2} + \ldots \nonumber\\
T(t)\,\ket{\Psi_3} &=& c_0 e^{-t\epsilon_0} \ket{\phi_0} + c_1 e^{-t\epsilon_1} \ket{\phi_1}
+ c_2 e^{-t\epsilon_2} \ket{\phi_2} + \ldots \nonumber\\
T(t)\,\ket{\Psi_4} &=& d_0 e^{-t\epsilon_0} \ket{\phi_0} + d_1 e^{-t\epsilon_1} \ket{\phi_1}
+ d_2 e^{-t\epsilon_2} \ket{\phi_2} + \ldots
\label{twostatedep} 
\end{eqnarray}
The reason it is convenient to make an orthogonal transformation on the states
$\ket{\Psi_i}$ is that in the $t \rightarrow \infty$ limit those states in
Eq.\ref{twostatedep} for which the coefficient of $\ket{\phi_0}$ is nonvanishing will,
up to a normalization factor, contract onto the same state $\ket{\phi_0}$, rendering
$[[T(t)\,H\,T(t)]]$ and $[[T^2(t)]]$ singular. By multiplying $[[T(t) H T(t)]]$
by the factors of $[[T(t)^2]]^{-1/2}$ we correct for this situation, but it is not
at all obvious why or how this works in the original tensor product basis.
  
To avoid this problem with the large $t$ limit we change
basis, defining states $\ket{\chi_1}$ to $\ket{\chi_4}$, which are orthonormal linear
combinations of the states $\ket{\Psi_1}$ to $\ket{\Psi_4}$, having the property that
each state contracts onto a distinct eigenstate of the block Hamiltonian; i.e., 
\begin{eqnarray}
\ket{\chi_1} &=& \alpha_0 \ket{\phi_0} + \alpha_1 \ket{\phi_1} + \alpha_2 \ket{\phi_2} +
\alpha_3 \ket{\phi_3} + \ldots \nonumber\\
\ket{\chi_2} &=& \phantom{ \alpha_0 \ket{\phi_0} +\  } \beta_1 \ket{\phi_1}
+ \beta_2 \ket{\phi_2}
+ \beta_3 \ket{\phi_3} + \nonumber\ldots\\
\ket{\chi_3} &=& \phantom{\alpha_0 \ket{\phi_0} + \alpha_1 \ket{\phi_1} + \ } \gamma_2 \ket{\phi_2} +
\gamma_3 \ket{\phi_3} + \ldots \nonumber\\
\ket{\chi_4} &=& \phantom{\alpha_0 \ket{\phi_0} + \alpha_1 \ket{\phi_1} + \alpha_2 \ket{\phi_2} + \  }
\delta_3 \ket{\phi_3} + \ldots
\label{twostateexptwo} 
\end{eqnarray}
Applying $T(t)$ to these basis states yields 
\begin{eqnarray}
T(t)\,\ket{\chi_1} &=& \alpha_0 e^{-t\epsilon_0} \ket{\phi_0} + \alpha_1 e^{-t\epsilon_1} \ket{\phi_1}
+ \alpha_2 e^{-t\epsilon_2} \ket{\phi_2} +  \alpha_3 e^{-t\epsilon_3} \ket{\phi_3} + \ldots \nonumber\\
T(t)\,\ket{\chi_2} &=& \phantom{ \alpha_0 e^{-t\epsilon_0} \ket{\phi_0}
+\  } \beta_1 e^{-t\epsilon_1} \ket{\phi_1} + \beta_2 e^{-t\epsilon_2} \ket{\phi_2}
+ \beta_3 e^{-t\epsilon_3} \ket{\phi_3} + \nonumber\ldots\\
T(t)\,\ket{\chi_3} &=& \phantom{\alpha_0 e^{-t\epsilon_0} \ket{\phi_0}
+ \alpha_1 e^{-t\epsilon_1} \ket{\phi_1} + \ } \gamma_2 e^{-t\epsilon_2}  \ket{\phi_2} +
\gamma_3 e^{-t\epsilon_3} \ket{\phi_3} + \ldots \nonumber\\
T(t)\,\ket{\chi_4} &=& \phantom{\alpha_0 e^{-t\epsilon_0} \ket{\phi_0}
+ \alpha_1 e^{-t\epsilon_1} \ket{\phi_1} + \alpha_2 e^{-t\epsilon_2} \ket{\phi_2} + \  }
\delta_3 e^{-t\epsilon_3} \ket{\phi_3} + \ldots
\label{Tstates}
\end{eqnarray}
from which it is clear that, up to a normalization factor, each of the states $\ket{\chi_\alpha}$
contracts onto a different eigenstate of the multi-block Hamiltonian.  Furthermore, we have
\begin{eqnarray}
H\,T(t)\,\ket{\chi_1} &=& \alpha_0 \epsilon_0 e^{-t\epsilon_0} \ket{\phi_0} + \alpha_1 \epsilon_1 e^{-t\epsilon_1} \ket{\phi_1}
+ \alpha_2 \epsilon_2 e^{-t\epsilon_2} \ket{\phi_2} +  \alpha_3 \epsilon_3 e^{-t\epsilon_3} \ket{\phi_3} + \ldots \nonumber\\
H\,T(t)\,\ket{\chi_2} &=& \phantom{ \alpha_0 \epsilon_0 e^{-t\epsilon_0} \ket{\phi_0}
+\  } \beta_1 \epsilon_1 e^{-t\epsilon_1} \ket{\phi_1} + \beta_2 \epsilon_2 e^{-t\epsilon_2} \ket{\phi_2}
+ \beta_3 \epsilon_3 e^{-t\epsilon_3} \ket{\phi_3} + \nonumber\ldots\\
H\,T(t)\,\ket{\chi_3} &=& \phantom{\alpha_0 \epsilon_0 e^{-t\epsilon_0} \ket{\phi_0}
+ \alpha_1 \epsilon_1 e^{-t\epsilon_1} \ket{\phi_1} + \ } \gamma_2 \epsilon_2 e^{-t\epsilon_2}  \ket{\phi_2} +
\gamma_3 \epsilon_3 e^{-t\epsilon_3} \ket{\phi_3} + \ldots \nonumber\\
H\,T(t)\,\ket{\chi_4} &=& \phantom{\alpha_0 \epsilon_0 e^{-t\epsilon_0} \ket{\phi_0}
+ \alpha_1 \epsilon_1 e^{-t\epsilon_1} \ket{\phi_1} + \alpha_2 \epsilon_2 e^{-t\epsilon_2} \ket{\phi_2} + \  }
\delta_3 \epsilon_3 e^{-t\epsilon_3} \ket{\phi_3} + \ldots
\label{HTstates}
\end{eqnarray}
Given Eq.\ref{Tstates} and Eq.\ref{HTstates} one can easily analyze the $t \rightarrow \infty$
behavior of Eq.\ref{hamtinf}.  To get a feeling for the way in which this works let us
simplify the discussion by, for the moment, assuming that $[[T^2(t)]]$ and
$[[T(t) H T(t)]]$ are $2\times 2$ matrices obtained by sandwiching these operators
between the states $\ket{\chi_1}$ and $\ket{\chi_2}$.
In this case we have
\begin{eqnarray}
[[T^2(t)]] &=& \left(
\begin{array}{c c }
	|\alpha_0|^2\, e^{-2\epsilon_0 t}+ \ldots
	& \alpha_1^\ast \,\beta_1\, e^{-2\epsilon_1 t }\nonumber \\
	\alpha_1\,\beta_1^\ast\, e^{-2\epsilon_1 t }
	& |\beta_1|^2\,e^{-2\epsilon_1 t}
\end{array} \right)
\end{eqnarray}
\begin{eqnarray}							
[[T(t) H T(t)]]	&=& \left(
	\begin{array}{c c }
          |\alpha_0|^2\,\epsilon_0\, e^{-2\epsilon_0 t} &
	  \alpha_1^\ast \,\beta_1\,\epsilon_1\, e^{-2\epsilon_1 t }\nonumber \\
	  \alpha_1\,\beta_1^\ast\,\epsilon_1 e^{-2\epsilon_1 t }
	  & |\beta_1|^2\,\epsilon_1\,e^{-2\epsilon_1 t}
	\end{array} \right)
\end{eqnarray}

In general $[[T^2(t)]]$ is a matrix of scalar products and its eigenvalues
are guaranteed to be positive, so it is guaranteed to have an inverse square root.
On can explicitly construct the inverse square root by writing $[[T^2(t)]]$ 
as
\begin{equation}
[[T^2(t)]] = U(t)^\dag\, D(t)\, U(t)
\end{equation} 
where $D(t)$ is a diagonal matrix whose entries are the eigenvalues of $[[T^2(t)]]$ and
$U(t)$ is the matrix whose columns are the normalized eigenvectors corresponding to
those eigenvalues.  Given this decomposition 
\begin{equation}
[[T^2(t)]]^{-1/2} = U(t)^\dag\, D(t)^{-1/2}\, U(t)
\end{equation}
Note, since $D(t)$ is diagonal, $D(t)^{-1/2}$ is also a diagonal matrix
whose entries are the inverse square roots of the corresponding entries in $D(t)$.

Fortunately, since all we really need is the behavior of the product
in Eq.\ref{hamtinf} as $t$ gets large, we don't have to do all this work.
It suffices to define a $t$-dependent rescaling of $\ket{\chi_1}$ and $\ket{\chi_2}$
which guarantees that each state converges, as $t \rightarrow \infty$, to the
lowest lying eigenstate of the block Hamiltonian which appears in its expansion
in terms of multi-block eigenstates.  More specifically, multiplying
$\ket{\chi_1}$ by the factor $e^{\epsilon_0\,t}/\alpha_0$ and
$\ket{\chi_2}$ by ${\epsilon_1\,t}/\beta_1$ yields the result
\begin{eqnarray}
\ket{\chi'_1(t)} &=& T(t)\,\frac{e^{\epsilon_0\,t}}{\alpha_0}\,\ket{\chi_1} =
\ket{\phi_0} + \frac{\alpha_1}{\alpha_0}\,e^{-(\epsilon_1-\epsilon_0)\,t} \ket{\phi_1}
+ \ldots
\nonumber \\ 
\ket{\chi'_2(t)} &=& T(t)\,\frac{e^{\epsilon_1\,t}}{\beta_1}\,\ket{\chi_2} =
\ket{\phi_1} + \ldots
\end{eqnarray}
from which it follows that in the limit of large $t$,
\begin{eqnarray}
[[T^2(t)]] &=& \left(
	\begin{array}{c c }
	   1 + \ldots
	   & \frac{\alpha_1^\ast}{\alpha_0^\ast}\, e^{-(\epsilon_1 -\epsilon_0)\,t }\nonumber \\
	   \frac{\alpha_1}{\alpha_0}\, e^{-(\epsilon_1-\epsilon_0)\,t }
	   & 1 + \dots
	\end{array} \right)
\end{eqnarray}
\begin{eqnarray}							
[[T(t) H T(t)]]	&=& \left(
	\begin{array}{c c }
	   \epsilon_0 + \ldots &
	   \frac{\alpha_1^\ast}{\alpha_0^\ast}\,\epsilon_1\,
	    e^{-(\epsilon_1-\epsilon_0)\, t }\nonumber \\
	   \frac{\alpha_1}{\alpha_0}\,\epsilon_1\,
	   e^{-(\epsilon_1-\epsilon_0)\, t }
	   & \epsilon_1 + \ldots
	\end{array} \right)
\end{eqnarray}
which establishes the theorem for this $2\times 2$ case.  It should be clear
that the same sort of rescaling of $\ket{\chi_1}$ to $\ket{\chi_4}$
will establish the result for the real $4\times 4$ case.  It is important to reiterate
that the construction the matrix of eigenvalues $D$ and the construction of the
orthogonal transformation $R$ is done directly from a knowledge of the eigenvalues
of the block Hamiltonian and the expansion of the tensor product states in
eigenstates of the block Hamiltonian; at no point is it necessary to deal with
$[[T(t)]]$ for finite values of $t$.  This means that when dealing with large
blocks and many eigenstates, techniques such as the Lanczos method, which finds only
the relevant lowest lying eigenstates starting from the tensor product states,
can greatly reduce the computer resources needed to solve the problem.

\section{Free-Field Preliminaries}

To set up the computation presented in the next section requires some notation.
The system under discussion is a $1+1$--dimensional
Hamiltonian lattice theory.  The fermion field is taken to be a two
component operator $\psi_\alpha(j)$ with $\alpha = 1,2$.  The Hamiltonian has the form
\begin{equation}
	H = \sum_{j_1,j_2} \frac{i}{2}\, \delta^{'}(j_1-j_2)\,\psi^{\dag}_{j_1} \sigma_3 \psi_{j_2}
\end{equation}
where $\delta^{'}$ is a general hopping term having the property
\begin{equation}
	\delta^{'}(j_1-j_2) = -\delta^{'}(j_2-j_1) .
\end{equation}
and $\sigma_3$ is the $2\times 2$ matrix
\begin{equation}
	\pmatrix{ 1 & \phantom{-}0 \cr
			  0 & -1\cr}
\end{equation}

As is customary I introduce creation and annihilation operators
$b_j , b^{\dag}_j, d_j , d^{\dag}_j$ and define the 
nearest neighbor derivative, $\delta^{'}(j)$, as follows:
\begin{eqnarray}
	\psi(j) &=& b_j \pmatrix{1 \cr 0} + d_j^{\dag}\pmatrix{0 \cr 1} \nonumber \\
	\delta^{'}(1) &=& -\delta^{'}(-1) = 1 \nonumber \\
	\delta^{'}(j) &=& 0 \quad {\rm if}\ j \ne 1
\end{eqnarray}
Making these substitutions the Hamiltonian takes the form
\begin{equation}
\label{nnhamiltonian}
	H = \sum_j \frac{i}{2} (b^{\dag}_{j+1} b_j - b^{\dag}_j b_{j+1}) - \frac{i}{2} (d^{\dag}_{j+1} d_j
	- d^{\dag}_j d_{j+1} ) 
\end{equation}
In terms of these operators the total electric and axial-charge are defined by
\begin{eqnarray}
	Q &=& \sum_j ( b_j^{\dag} b_j - d_j^{\dag} d_j ) \nonumber \\
	Q^5 &=& \sum_j ( b_j^{\dag} b_j + d_j^{\dag} d_j - 1 )
\end{eqnarray}
For each site $j$, there are two electrically neutral states.  The first is the
state $\ket{0_j}$ which is annihilated by $b_j$ and $d_j$.  The other chargeless
(i.e. locally gauge-invariant) state is $\ket{\pm_j}=b_j^{\dag}d_j^{\dag}\ket{0_j}$.
The projection operator onto the space of {\it retained states\/} is defined
in terms of these states to by the product
\begin{eqnarray}
	P &=& \prod_j P_j \nonumber \\
	P_j &=& \ket{0_j}\bra{0_j} + \ket{\pm_j}\bra{\pm_j} .
\end{eqnarray}
	
Fourier transforming 
\begin{eqnarray}
	b_k &=& \sum_j e^{-ikj} b_j ;\qquad	b_k^{\dag} = \sum_j e^{ikj} b_j^{\dag} \nonumber \\
	d_k &=& \sum_j e^{-ikj} d_j ; \qquad d_k^{\dag} = \sum_j e^{ikj} d_j^{\dag} 
\end{eqnarray}
the Hamiltonian in Eq.\ref{nnhamiltonian} becomes
\begin{equation}
\label{fthamiltonian}
	H = \frac{1}{2\pi} \int_{-\pi}^{\pi} dk  \sin(k)\, ( b^{\dag}_k b_k - d^{\dag}_k d_k ) 
\end{equation}
It follows from Eq.\ref{fthamiltonian} that the ground-state of the theory
is the state obtained by filling the negative energy sea; i.e.,
\begin{equation}
	\ket{\rm vac} = \prod_{k \ge 0 } b^{\dag}_{-k} d^{\dag}_k \ket{0}
\end{equation}
where $\ket{0}$ is the state annihilated by all the $b_k$'s and $d_k$'s (
or equivalently, the $b_j$'s and $d_j$'s).

The only additional free-fermion formula which I need has to do with
diagonalizing Eq.\ref{nnhamiltonian} for a finite lattice where $j=1..N$.
As with all such quadratic Hamiltonians one only has to diagonalize
the $N\times N$ matrix $K_{j_1,j_2} = i/2\,\delta^{'}(j_1-j_2)$, which can
be done exactly for any value of $N$.  If $N$ is even the eigenvalues
of $K$ look like the eigenvalues of the infinite volume problem in that
\begin{equation}
\label{eigenvals}
	\epsilon (k) = \sin(k_p)
\end{equation}
where $k_p = \pi p /2(N+1) $ for $p$ any odd integer such that $-N \le p \le N$.
For the case of odd $N$, the eigenvalues are once given by
Eq.\ref{eigenvals}, however in this case the values of $k_p$ are given
by $k_p = \pi p /(N+1)$ where now $p$ is any integer such that $-(N+1)/2 < p < (N+1)/2$.
Note that in both cases the range of $k$ is $ -\pi/2 < k < \pi/2 $ in distinction
to the case of a periodic or infinite lattice.  Although the issue of how doubling
works in the finite volume open boundary condition case is interesting,
time does not permit going into it here.  Suffice it to say
that even though the eigenvalues are given by Eq.\ref{eigenvals}
and undoubled, nevertheless the theory has both left and right movers for
each component of $\psi$ and is not a chiral theory.
The last piece of information needed in order to carry out the full
computation is the formula for the eigenvectors of $K$, which for all values
of $N$ are given by
\begin{equation}
	u_{k_p}(j) = \frac{1}{\sqrt{2(N+2)}}\,( e^{-i k_p j} - (-1)^j e^{i k_p j} )
\end{equation}

A final fact concerning the nearest-neighbor free-fermion theory is that
it possesses a global $SU(2)$ symmetry whose generators are given by
\begin{eqnarray}
\label{globalsutwo}
	S_\alpha &=& \sum_j  \frac{1}{2} \psi^{\dag}_j \sigma_3^j \sigma_\alpha \sigma_3^j \psi_j \nonumber \\
			S_{+}	&=&\sum_j (-1)^j b^{\dag}_j d^{\dag}_j \nonumber \\
			\label{splus}
			S_{-} &=& \sum_j (-1)^j d_j b_j \nonumber \\
			\label{sminus}
			S_3  &=& \frac{1}{2} ( b^{\dag}_j b_j + d^{\dag}_j d_j -1 ) 
\end{eqnarray}
This symmetry is very useful for checking the results of the necessary
finite volume calculations.

\section{The Computation}

To compute the connected range-1 term $H^{\rm conn}(j,1)$ one
diagonalizes the single-site Hamiltonian and selects a set of retained states
from among its lowest lying eigenstates.
Since there are no range-1 terms in Eq.\ref{nnhamiltonian}
the single-site Hamiltonian is zero and so, independent of the choice
of retained states $h^{\rm conn}(j,1) = 0$.  As noted earlier,
I will retain the locally gauge-invariant states
\begin{equation}
	\ket{0_j} ; \qquad \ket{\pm_j} = b^{\dag}_j d^{\dag}_j \, \ket{0_j} .
\end{equation}
which form a spin-1/2 doublet with respect to the global $SU(2)$
defined in Eq.\ref{globalsutwo}.

The first non-trivial connected operator which contributes to $H^{\rm ren}$
is the range-2 operator $h^{\rm conn}(j,2)$.  The four retained
states associated with a connected two-site block are
\begin{equation} 
	\ket{0 0} ; \quad \ket{\pm 0} ; \quad \ket{ 0, \pm} ; \quad \ket{\pm \pm}
\end{equation} 
which decompose into a spin-1 and spin-0 multiplet under the global
$SU(2)$.  Since each of the states of definite spin and third-component of spin
must contract onto unique eigenstates of the two-site problem, 
changing from the tensor product states to the spin basis amounts 
to finding the rotation matrix $R$.  The combinations of definite
total spin and 3-component of spin are:
\begin{eqnarray}
	S_{\rm tot}^2 = 1 &\qquad & S_3 = -1 \qquad \ket{0 0} \nonumber \\
	S_{\rm tot}^2 = 1 &\qquad & S_3 = \phantom{-}0 \qquad \frac{1}{\sqrt{2}} (\ket{\pm 0} -
	\ket{0 \pm} ) \nonumber \\
	S_{\rm tot}^2 = 1 &\qquad & S_3 = \phantom{-}1 \qquad \ket{\pm \pm}\nonumber \\
	S_{\rm tot}^0 = 1 &\qquad & S_3 = \phantom{-}0 \qquad \frac{1}{\sqrt{2}}
	(\ket{\pm 0} + \ket{0 \pm} ) 
\end{eqnarray}
Note the unusual minus sign in the spin-1, $S_3=0$ state, 
comes from the alternating minus sign which appears
in the definition of $S_{+}$ and $S_{-}$ in Eqs.\ref{splus} and Eq.\ref{sminus}.  

To complete the computation of $h^{\rm conn}(j,2)$ we only need to find the
eigenvalues of the full two-site Hamiltonian which correspond to the lowest
lying spin-0 and spin-1 states.  Since the Hamiltonian, Eq.\ref{nnhamiltonian},
only has terms which absorb a particle or antiparticle at site $j$ and create
one at $j \pm 1$ it follows that $\ket{0 0 }$, which is annihilated by the
absorption operators, has zero energy.  The same is true of the state $\ket{\pm \pm}$,
which is not annihilated by the absorption operators but which has no room to move
a particle or antipartile to a neighboring site.  It follows from the $SU(2)$ symmetry that
the state $\frac{1}{\sqrt{2}} (\ket{\pm 0} - \ket{0 \pm})$ also has zero energy.
The only eigenvalue to be determined is that associated with the lowest spin-0
state in the sector which has one particle and one antiparticle.  For the case
of two sites the two allowed values for the momenta are $k = \pm \pi/6 $
and $\epsilon(k) = \pm \sin(\pi/6) = \pm 1/2 $; thus the lowest energy charge zero
eigenstate of the two-site problem is $ b^{\dag}_{-\pi/6} d^{\dag}_{\pi/6} \ket{0,0}$,
having an energy $E_2 = -1$.  It follows from this that, in the total spin basis
\begin{equation}
\label{eqna}
	H_2^{\rm ren}(i, i+1) = R \, \left(\begin{array}{c c c c}
	 0 & 0 & 0 & \phantom{-} 0 \\
	 0 & 0 & 0 & \phantom{-} 0 \\
	 0 & 0 & 0 & \phantom{-} 0 \\
	 0 & 0 & 0 & -1 
	\end{array}\right) \, R^{\dag}
\end{equation}
where $R$ is the rotation matrix which transformed the original tensor products
into the total spin basis.  This $4\times 4$ matrix can always be rewritten
in terms of the sixteen matrices $M_{\alpha \beta} = \sigma_\alpha(i) \sigma_\beta(i+1)$
where $\alpha,\beta = 0 \ldots 3$ and $\sigma_0$ is the unit matrix and
$\sigma_1$, $\sigma_2$ and $\sigma_3$ are the Pauli spin matrices.  In particular
if we write
\begin{equation}
\label{sigmaab}
 H_2^{\rm ren}(i,i+1) = \sum_{\alpha, \beta = 0}^4 c_{\alpha \beta} M_{\alpha \beta}
\end{equation}
then it follows from the $SU(2)$-symmetry and Eq.\ref{eqna} that, in the original
tensor product basis, $H_2^{\rm ren}(i,i+1)$ has the form
\begin{equation}
	h^{\rm conn}(2,i) = H_2^{\rm ren}(i,i+1) = -\frac{1}{4} 1 + \vec{s}(i) \cdot \vec{s}(i+1)
\end{equation}
since $h^{\rm conn}(1,i)=0$.  Thus, up to range-2 terms, the full renormalized Hamiltonian is
\begin{equation}
\label{rotatHam}
	H^{\rm ren} =  -\frac{1}{4} V  + \sum_i \vec{s}(i) \cdot \vec{s}(i+1)
\end{equation}
Actually, this form of the renormalized Hamiltonian gives the correct spectrum,
but it is not what one obtains directly by taking traces of the form
${\rm tr}(s_x(i) s_x(i+1) H_2^{\rm ren}(i,i+1))$
as indicated in Eq.\ref{sigmaab}.  The nonvanishing traces are
\begin{eqnarray}
\label{unrotated}
	\frac{1}{4} {\rm tr}(\sigma_1(i) \sigma_1(i+1) H_2^{\rm ren}(i,i+1)) &=& -\frac{1}{4} \nonumber \\
	\frac{1}{4} {\rm tr}(\sigma_2(i) \sigma_2(i+1) H_2^{\rm ren}(i,i+1)) &=& -\frac{1}{4} \nonumber \\
	\frac{1}{4} {\rm tr}(\sigma_3(i) \sigma_3(i+1) H_2^{\rm ren}(i,i+1)) &=& \phantom{-}\frac{1}{4} 
\end{eqnarray}
and so the form of $H_2^{\rm ren}$ directly obtained from our definitions has the form
\begin{eqnarray}
	H_2^{\rm ren} &=& \sum_i H_2^{\rm ren}(i,i+1) \nonumber \\
	&=& -\frac{1}{4} V + \sum_i (-s_1(i)\,s_1(i+1)-s_2(i)\,s_2(i+1)+s_3(i)\,s_3(i+1)\nonumber \\
\end{eqnarray}
Note, however, that this Hamiltonian can be brought into the form of a
Heisenberg antiferromagnet by the rotation
\begin{equation}
	O = \prod_j e^{-i \pi s_3(2j+1)}
\end{equation}
which, for every odd lattice site maps $s_1(2j+1) \rightarrow -s_1(2j+1)$ and 
$s_2(2j+1) \rightarrow -s_2(2j+1)$ but leaves $s_3(2j+1)$ unchanged.  In the calculations
which follow we will see that, given the phases I have chosen for for the states $\ket{0_j}$ and
$\ket{\pm_j}$, the terms in $H_r^{\rm ren}$ will generically have the form shown in Eq.\ref{unrotated} 
and will be brought into standard form by application of $O$.

While the computation of $h^{\rm conn}(1,i)$ and $h^{\rm conn}(2,i)$ are very simple,
the calculation of $h^{\rm conn}(n,i)$ for $n \ge 3$ requires more work.  Fortunately,
it is easy to automate this task using MapleV and the computation of all terms out to
and including $h^{\rm conn}(5,i)$ takes less than two minutes on a desktop computer.
I will now describe the process for the case $h^{\rm conn}(3,i)$, since the
general computation proceeds along the same lines.

The allowed momenta for a three-site sublattice are $k = -\pi/4, 0, \pi/4$;
the corresponding particle energies are $\sin(-\pi/4)=-1/\sqrt{2}, \sin(0)=0 ,\sin(\pi/4)= 1/\sqrt{2}$; and
the antiparticle energies are $ -\sin(-\pi/4) = 1/\sqrt{2}, -\sin(0)=0 , -\sin(\pi/4)= -1/\sqrt{2} $. 
As the Hamiltonian contains only hopping terms, the number of particles and antiparticles
are separately conserved; thus, of the eight possible tensor product states (which define the
set of retained states) the states $\ket{000}$ and $\ket{\pm \pm \pm}$ are eigenstates
of the three-site Hamiltonian of energy zero.  Furthermore, of the six-remaining states,
the three two-particle states $\ket{\pm 0 0}$, $\ket{0 \pm 0}$ and $\ket{0 0 \pm}$
and the three four-particle states $\ket{\pm \pm 0}$, $\ket{\pm 0 \pm}$ and $\ket{0 \pm \pm}$
can be treated separately, since when expanded in terms of a complete set of eigenstates
for the three-site problem they have no states in common.

The nine possible particle-antiparticle eigenstates of the three site problem
are
\begin{equation}
 \ket{k,l} = b^{\dag}_k d^{\dag}_l \ket{0 0 0 }
\end{equation}
with eigenenergy $\epsilon(k,l) = \sin(k) - \sin(l)$.
Expanding the three retained states in terms of these eigenstates is straightforward
since
\begin{equation}
	b^{\dag}_j d^{\dag}_j \ket{0 0 0} = \sum_{k,l} u_k(j) u_l(j)
		b^{\dag}_k d^{\dag}_l \ket{0 0 0}
\end{equation}
thus, the coefficient $\alpha_{k l}$ of the state $b^{\dag}_k d^{\dag}_l$ is
just
\begin{equation}
	\alpha_j(k,l) = \sum_{k,l} u_k(j) u_l(j)
\end{equation}

Given these formulae we can now discuss how to compute the matrices
$R$,$H_{\rm diag}$ and $R^{\dag}$ from the overlap matrix, which gives the expansion
of the retained states in terms of the eigenstates of the three-site Hamiltonian.
The eigenenergies of the one particle states ordered by energy are:
\begin{equation}
\begin{array}{l c l c l}
\epsilon(-\pi/4,\pi/4) = -\sqrt{2} & , & \epsilon(-\pi/4,0) = -1/\sqrt{2} & , & 
\epsilon(0,\pi/4) = -1/\sqrt{2} \\
\epsilon(-\pi/4,-\pi/4) = 0 & , & \epsilon(0, 0) = 0 & , & \epsilon(\pi/4,\pi/4) = 0 \\ 
\epsilon(\pi/4,0) = 1/\sqrt{2} & , & \epsilon(0,-\pi/4) = -1/\sqrt{2} & , & 
\epsilon(\pi/4,-\pi/4) = \sqrt{2}
\end{array}
\end{equation}
The first overlap matrix giving the expansion of the retained states in terms
of these eigenstates is
\begin{equation}
O_v = 
\begin{array}{c} \bra{\pm\, 0\, 0} \\ \bra{0 \pm 0} \\ \bra{0\, 0\, \pm}\end{array}
\left(\begin{array}{c c c c c c c c c}
.25 & .35 & .35 & .25 & .25 & .5 & .35 & \cdots & \cdots \\
.50 & 0 & 0 & -.5 & -.5 &  0 & 0 & \cdots & \cdots \\
.25 &  -.35 & -.35 & .25 & .25 & .5 & -.35 & \cdots & \cdots 
\end{array}
\right)
\end{equation}
Focusing on the first column of this matrix it is easy to construct a $3 \times 3$-rotation
matrix, $R_1$ which takes the first column into a vector with one nonvanishing
component; i.e.
\begin{equation}
R_1 O_v = \left(\begin{array}{c c c c c c c c c}
.61 &  0 & 0 & -.20 & -.20 & .41 & 0 & \cdots & \cdots \\
0 & -.39 & -.39 &  -.37 & -.37 & -.38 & -.39 & \cdots & \cdots \\
0 &  -.32  & -.32 & .45 & .45 &  .45 & -.32 & \cdots & \cdots
\end{array} \right)
\end{equation}
The fact that the first column has one nonvanishing entry means that it is
the only state which has an overlap with the three-site groundstate, which is
the first objective we wished to achieve.  Now, the fact that the last two
entries in the second column are nonvanishing means that we are still in a
situation where both the second and third rotated retained states contract
onto the first excited state of the three-site problem.  Clearly we can
now perform a rotation $R_2$ which is the identity on the first rotated
state and rotates the second two states so that in the new rotated basis
only the second element of the second column of $R_2 R_2 O_v$ is nonvanishing;
i.e.,
\begin{equation}
R_2 R_1 O_v = \left(\begin{array}{c c c c c c c c c}
.61 & 0  & 0 & -.20 & -.20 & .41 & 0 & \cdots & \cdots \\
0 & .5 & .5 & 0 & 0 & 0 & .5 & \cdots & \cdots \\
0 & 0 & 0 & -.58 & -.58 & -.58 & 0 & \cdots & \cdots  
\end{array}
\right)
\end{equation}
Now we have finally achieved our original goal, each of the rotated states
will contract onto a unique eigenstate of the three-site Hamiltonian;
i.e., the first rotated state contracts onto the state
$b^{\dag}_{-\pi/4} d^{\dag}_{\pi/4} \ket{0 0 0}$, the second state contracts onto
$(b^{dag}_{-\pi/4} d^{\dag}_0 + b^{\dag}_0 d^{\dag}_{\pi/4})\ket{0 0 0}/\sqrt{2} $
and the third state contracts onto $(b^{\dag}_{-\pi/4} d^{\dag}_{-\pi/4} +
b^{\dag}_0 d^{\dag}_0 + b^{\dag}_{\pi/4} d^{\dag}_{\pi/4})\ket{0 0 0}/\sqrt{3} $.
Thus, in this new rotated basis the diagonal Hamiltonian in the three-site one pair
sector is
\begin{equation}
H_{\rm D}(\rm 1-pair) = \left( 
{\begin{array}{ccr}
-\sqrt{2} & 0 & 0 \\
0 & -1/\sqrt{2} & 0 \\
0 & 0 & 0
\end{array}}
 \right)
\end{equation} 
Given the explicit form of the rotation matrix $R = R_2 R_1$
we can compute the Hamiltonian in the original tensor product basis to be
\begin{equation}
H_3^{\rm ren}(\rm 1-pair) = R H_{\rm D}(\rm 1-pair) R^{\dag} =
\left({\begin{array}{ccc}
-.589 & -.471 & .118 \\
-.471 & -.943 & -.471 \\
.118 & -.471 & -.589
\end{array}}
 \right)
\end{equation}

A similar calculation can be done for the two-pair sector of the three-site
problem.  Here the three retained states are $\ket{\pm \pm 0}$, $\ket{\pm 0 \pm}$ and
$\ket{0 \pm \pm}$.  Obviously, as in the one-pair case, these states are of the
generic form
\begin{equation}
	\ket{j_1, j_2} = b^{\dag}_{j_1} d^{\dag}_{j_1} b^{\dag}_{j_2} d^{\dag}_{j_2} \ket{0 0 0}
\end{equation}
and can be directly rewritten in terms of the eigenstates of the three-site
Hamiltonian using the operators $b^{\dag}_k$ and $d^{\dag}_l$ as
\begin{eqnarray}
\label{expansiontwo}
    \ket{j_1, j_2} &=& \sum_{k_1, k_2, l_1, l_2} u_{k_1}(j_1) u_{l_1}(j_1) u_{k_2}(j_2)
		u_{l_2}(j_2) b^{\dag}_{k_1} d^{\dag}_{l_1} b^{\dag}_{k_2} d^{\dag}_{l_2}
		\ket{0 0 0} \nonumber \\
		&=& \sum_{k_1 < k_2} \sum_{l_1<l_2} 
		(u_{k_1}(j_1) u_{k_2}(j_2) - u_{k_2}(j_1) u_{k_1}(j_2) ) \times\nonumber \\
		&\phantom{=}& \sum_{k_1 < k_2} \sum_{l_1<l_2} (u_{l_1}(j_1) u_{l_2}(j_2) - u_{l_2}(j_1) u_{l_1}(j_2) )
		\ket{k_1, k_2, l_1, l_2} 	
\end{eqnarray}
where by $\ket{k_1, k_2, l_1, l_2}$ we denote the state $b^{\dag}_{k_1} d^{\dag}_{l_1} b^{\dag}_{k_2}
d^{\dag}_{l_2} \ket{0 0 0}$ for the particular ordering $k_1 < k_2$ and $l_1 < l_2$.
With this choice of a complete set of basis vectors it is easy
to read off the overlap matrix; it is not surprising, given the $SU(2)$ symmetry of the nearest
neighbor problem, that it is identical to the overlap matrix of the one-pair sector and
so the same procedure can be applied to bring it to the desired form.

Note that the no-pair state $\ket{0 0 0}$ and the three-pair state $\ket{\pm \pm \pm}$
are each exact eigenstates of the three-site problem with eigenvalue zero so
\begin{equation}
H_3^{\rm ren}(\rm 0-pair)(i, i+1, i+2) = H_3^{\rm ren}(\rm 3-pair)(i, i+1, i+2) = 0
\end{equation}

Given these results it is now trivial to expand $H_3^{\rm ren}$ as
\begin{equation}
\label{hthreeexp}
	H_3^{\rm ren}(i, i+1, i+2) = \sum_{\alpha \beta \gamma} c_{\alpha \beta \gamma}
	\sigma_{\alpha}(i) \sigma_{\beta}(i+1) \sigma_\gamma(i+2)
\end{equation}
by taking traces;  thus, for example, the coefficient of the identity operator is
\begin{equation}
c_{0 0 0} = \frac{1}{8} {\rm tr} (\sigma_\alpha(i) \sigma_\beta(i+1)
	\sigma_\gamma(i+2) H_3^{\rm ren}(i, i+1, i+2)) .
\end{equation}
Proceeding in this way we obtain, for the case of three sites,
\begin{equation}
  c_{0 0 0}^{\rm 3-sites} =  -.53033\ldots
\end{equation}
To convert this to a connected contribution we have to subtract the two ways of embedding
the two-site problem into the three-site problem, obtaining
\begin{eqnarray}
	c_{0 0 0}^{\rm 3-sites conn} &=& c_{0 0 0}^{\rm 3-sites} - 2 c_{0 0}^{\rm 2-sites conn} \nonumber \\
	&=& -.53033\ldots -2 (-1/4) = -.03033\ldots
\end{eqnarray} 
which shows that the range-3 connected contribution to the ground-state energy density
is quite small in comparison to the range-2 contribution.  Similarly we can compute the
coefficients of the other operators which can appear in Eq.\ref{hthreeexp}.  Thus, we have
\begin{eqnarray}
c_{1 1 0}^{\rm 3-sites} &=& =c_{2 2 0}^{\rm 3-sites} = -c_{3 3 0}^{\rm 3-sites} = -.23570\ldots \nonumber \\
c_{0 1 1}^{\rm 3-sites} &=& =c_{0 2 2}^{\rm 3-sites} = -c_{0 3 3}^{\rm 3-sites} = -.23570\ldots \nonumber \\
c_{1 0 1}^{\rm 3-sites} &=& =c_{2 0 2}^{\rm 3-sites} =  c_{3 0 3}^{\rm 3-sites} = -.05893\ldots 
\end{eqnarray}
and all other possible $c_{\alpha \beta \gamma}$ vanish.

As we already noted, for nearest neighbor spin-spin interactions, which is what
$c_{1 1 0}$, $c_{2 2 0}$ and $c_{3 3 0}$  parameterize, there should be a difference in sign between 
$c_{3 3 0}$ and both $c_{1 1 0}$ and $c_{2 2 0}$, which is the case; the same is true
for $c_{0 1 1}$, $c_{0 2 2}$ and $c_{0 3 3}$.  Note however, the next to nearest neighbor
coefficients $c_{1 0 1}$, $c_{2 0 2}$ and $c_{3 0 3}$ should have the same sign and they do.

To obtain the connected coefficients we recall that
\begin{equation}
 H_3^{\rm conn}(i, i+1, i+2) = H_3^{\rm ren}(i, i+1, i+2) - H_2^{\rm conn}(i, i+1) - H_2^{\rm conn}(i+1, i+2) .
\end{equation}
Using the explicit form for $H_3^{\rm ren}(i, i+1, i+2)$, $H_2^{\rm conn}(i, i+1)$ and
$H_2^{\rm conn}(i+1, i+2)$ and collecting like terms we obtain
\begin{eqnarray}
	c_{1 1 0}^{\rm conn} &=& c_{1 1 0}^{\rm 3-sites} - c_{1 1}^{\rm 2-sites} = 0.014297\ldots \nonumber \\
	c_{2 2 0}^{\rm conn} &=& c_{2 2 0}^{\rm 3-sites} - c_{2 2}^{\rm 2-sites} = 0.014297\ldots \nonumber \\
	c_{3 3 0}^{\rm conn} &=& c_{3 3 0}^{\rm 3-sites} - c_{3 3}^{\rm 2-sites} = -0.014297\ldots \nonumber  \\
	c_{0 1 1}^{\rm conn} &=& c_{0 1 1}^{\rm 3-sites} - c_{1 1}^{\rm 2-sites} = 0.014297\ldots \nonumber \\
	c_{0 2 2}^{\rm conn} &=& c_{0 2 2}^{\rm 3-sites} - c_{2 2}^{\rm 2-sites} = 0.014297\ldots \nonumber \\
	c_{0 3 3}^{\rm conn} &=& c_{0 3 3}^{\rm 3-sites} - c_{3 3}^{\rm 2-sites} = -0.014297\ldots \nonumber \\
	c_{1 0 1}^{\rm conn} &=&  0.05892\ldots \nonumber \\
	c_{2 0 2}^{\rm conn} &=&  0.05892\ldots \nonumber \\
	c_{3 0 3}^{\rm conn} &=&  0.05892\ldots 
\end{eqnarray}
all other coefficients vanish.

Thus, recalling the for $\alpha = 1,2,3$ $s_\alpha = \sigma_\alpha /2$, adding up the
range-2 and range-3 connected coefficients and rotating $s_1$ and $s_2$
to minus themselves on alternate sites, we can rewrite the range-3 renormalized Hamiltonian
as
\begin{equation}
	H^{\rm ren} = -0.28033 V + \sum_i \left[ 0.9428 \vec{s}(i)\cdot \vec{s}(i+1)
	+ .23570 \vec{s}(i) \cdot \vec{s}(i+2)	\right] 
\end{equation}
which has the advertised form of a frustrated Heisenberg antiferromagnet.
The important things to notice about the result of this calculation is that the range-3
corrections to the coefficients of operators which appeared at range-2 are small and that
the coefficients of the new operators which appear for the first time at range-3 are
typically smaller than the ones which appear at range-2.

This process can be carried out in the same way to compute $H_n^{\rm conn}$.  The result
out to and including range-5 contributions is
\begin{eqnarray}
\label{hrenfinal}
	H^{\rm ren} &=& -.31099\, V + \sum_i H_{\rm 2-body}(i) + \sum_i H_{\rm 4-body}(i) \nonumber \\
	H_{\rm 2-body}(i) &=& 0.80001\, \vec{s}(i)\cdot \vec{s}(i+1)
	+ 0.23492 \,\vec{s}(i) \cdot \vec{s}(i+2) - 0.01915\,  \vec{s}(i)\cdot\vec{s}(i+3) \nonumber \\
	H_{\rm 4-body} &=&  0.03559\, \vec{s}(i)\cdot \vec{s}(i+1)\  \vec{s}(i+2)\cdot\vec{s}(i+3)\nonumber \\
	&-& 0.08033 \, \vec{s}(i)\cdot\vec{s}(i+2)\  \vec{s}(i+1)\cdot\vec{s}(i+3)\nonumber \\
	&+& 0.03403 \, \vec{s}(i)\cdot\vec{s}(i+3) \  \vec{s}(i+1)\cdot\vec{s}(i+2) \nonumber \\
	&+& 0.02595\, \vec{s}(i)\cdot\vec{s}(i+1)\  \vec{s}(i+2)\cdot\vec{s}(i+4) \nonumber \\
	&+& 0.00339\, \vec{s}(i)\cdot\vec{s}(i+1)\  \vec{s}(i+3)\cdot\vec{s}(i+4) \nonumber \\
	&-& 0.01159 \, \vec{s}(i)\cdot\vec{s}(i+2) \   \vec{s}(i+1)\cdot\vec{s}(i+4) \nonumber \\
	&+& 0.05189 \, \vec{s}(i)\cdot\vec{s}(i+2) \  \vec{s}(i+3)\cdot\vec{s}(i+4) \nonumber \\
	&-& 0.03289 \, \vec{s}(i)\cdot\vec{s}(i+3) \  \vec{s}(i+1)\cdot\vec{s}(i+4) \nonumber \\
	&-& 0.01159 \, \vec{s}(i)\cdot\vec{s}(i+3) \  \vec{s}(i+2)\cdot\vec{s}(i+4) \nonumber \\
	&-& 0.00732 \, \vec{s}(i)\cdot\vec{s}(i+4) \  \vec{s}(i+1)\cdot\vec{s}(i+2) \nonumber \\
	&-& 0.03251 \, \vec{s}(i)\cdot\vec{s}(i+4) \  \vec{s}(i+1)\cdot\vec{s}(i+3) \nonumber \\
	&-& 0.00732 \, \vec{s}(i)\cdot\vec{s}(i+4) \  \vec{s}(i+2)\cdot\vec{s}(i+3) \nonumber \\	
\end{eqnarray}

The important fact to notice about all of these calculations is that at each stage
the tensor product states contract onto the lowest energy states of the corresponding
block Hamiltonian, thus proving that, if one computes out to infinite range,
the renormalized Hamiltonian will describe zero-charge sector of the free
system.  Furthermore, note that if one uses the same method to
calculate the two-point fermion antifermion correlation function CORE will guarantee that
its renormalized operator has the same matrix elements in the frustrated HAF ground-state
that the original operators had in the free-fermion vacuum state.

\section{Discussion}

Although, in principle, there are an infinite number of terms in $H^{\rm ren}$ Eq.\ref{hrenfinal}
shows that the coefficients of the longer range two-body terms, as well as the coefficients
of the four-body and higher terms drop off quickly; thus, as advertised in the introduction,
we see that the zero-charge sector of the free (doubled) fermion theory maps into a generalized
frustrated antiferromagnet. Frustrated, in this context, simply means that the coefficient of
the range-two and range-three one-body terms are both positive and the range-four term is negative.
To understand what is going on physically consider a N\'eel state in which the
spins are oriented $\ket{\ldots \uparrow \downarrow
\uparrow \downarrow \ldots}$.  It is clear, restricting attention to the nearest-neighbor
$\vec{s}(i)\cdot\vec{s}(i+1)$ terms, that the fact that the coefficient of these terms is
positive means that this orientation of the spins minimize this contribution to the
energy.  This, of course, implies that spins separated by two sites
should point in the same direction.  However, this maximizes instead of minimizes the
range-three spin-spin term since its coefficient is also positive.  Note also the the range-four
term also works against the range-two term since it wants spins separated by three sites to
point in the same direction, which is opposite to what one obtains from miminizing the
range-two term.  The presence of the four-body (and higher body) operators further confuses
the issue.  Since little is known about the physics of frustrated antiferromagnets, it is 
of some interest to know that this generalized class of theories are related to free fermion theories.
In fact, I will now argue that this is true for a non-trivial range of parameters.

To begin note that
it is easy to modify the free fermion derivative by adding hopping terms which jump three sites,
five sites, seven sites, etc., in such a way that the infinite volume kinetic energy has the general form
\begin{equation}
	\epsilon(k) = \sum_{m=1}^R a_r\, \sin((2r+1)k)
\end{equation}
This form of $\epsilon(k)$ is still symmetric about $k=\pi/2$ and exhibits the same sort of fermion
doubling as is exhibited by the nearest neighbor term.  In this case it is
a little more difficult to carry out the computations of the previous section, but they
go through in essentially the same way, except that the coefficients of the
operators in $H^{\rm ren}$ are changed.  It therefore follows that there is
more than one generalized frustrated antiferromagnet which is equivalent to a theory
of free fermions.  Since all of the various fermion derivatives can be defined so that
the slope of $\epsilon(k)$ is unity for small values of $k$, it follows that
all of these different theories must flow, under real-space renormalization group
transformations, to a single fixed-point generalized frustrated antiferromagnet.
This means that there is a nontrivial surface in the parameter space of these generalized
frustrated antiferromagnets which at low energies describe the same system of free massless fermions.

Having said this, it is important to point out that not all of these frustrated antiferromagnets
correspond to theories of free fermions.  As I pointed out in the introduction the same
mapping can be carried out for the case of an interacting gauge theory.  In fact,
the original motivation of this work was to do exactly this for lattice QCD
and analyze the weak coupling regime.  Obviously, constructing the mapping which takes the interacting
gauge theory into this class of generalized frustrated antiferromagnets is technically
more challenging, but it must exist.  From this it follows that for some choice
of couplings the generalized frustrated antiferromagnet corresponds not to
free fermions, but rather to a theory with fermions and gauge fields, neither of which is
apparent when one looks at the theory at the single-site or few-site level.

A final issue which needs to be addressed is how things work for higher dimension
and/or non-abelian gauge theories.  Clearly, for the case of doubled
free fermion theories in $3+1$-dimensions there are no new technical challenges; however,
the number of single-site gauge-invariant states which can be constructed is larger
since the particles and antiparticles come with two different spins.  As in the
$1+1$-dimensional case, these states form an irreducible representation of a symmetry
group, in this case a six-dimensional representation of $SU(4)$ and truncating to
this set of states, using CORE to compute the renormalized Hamiltonian, leads
to a generalized frustrated $SU(4)$ antiferromagnet (i.e., one where the $\vec{s}\cdot\vec{s}$
terms are replaced by $\vec{Q}\cdot\vec{Q}$ terms, where the $\vec{Q}$'s are now the
representation matrices for the six-dimensional representation of the generators of
$SU(4)$).  In this case using {\it generalized\/} to modify the phrase {\it frustrated
antiferromagnet\/} is even more appropriate since in this theory a new kind of
interaction term appears in the renormalized Hamiltonian.  This happens because
the six-dimensional representation of $SU(4)$ is not the fundamental representation
and so operators of the form $(\vec{Q}\cdot\vec{Q})^2$ can appear, and these operators
can significantly alter the landscape of possible phases which the theory can have.
A discussion of how such terms can modify the behavior of an antiferromagnetic theory
will be given in a forthcoming paper on the Haldane conjecture.\cite{Haldane}
Note, generally all terms permitted by the global symmetries of the problem
can and will appear.

As to the case of free quarks in $3+1$-dimensions, things are again
different in detail, but the general results are similar.  Here we are
interested in all the color singlet states which
can be formed from quarks and anti-quarks on a single site.  If one restricts to just
mesons, i.e. all states formed from color-singlet quark anti-quark states, then
the states on a single site fall into an irreducible representation of
$SU(12)$, which is a symmetry of the doubled theory.  Once again, if one truncates
to just this set of states the resulting renormalized Hamiltonian must be a generalized
frustrated $SU(12)$ antiferromagnet.  Once again generalized antiferromagnet
must be taken in the broadest possible sense since new terms of the form discussed
for the case of a single fermion in $3+1$-dimensions will appear, but now one will
also get new manybody terms related to the possibility of having Casimir
operators beyond the simplest quadratic Casimir operator.  Thus one should expect that
theories which correspond to quarks in interaction with color-gauge fields to have
a very rich structure.  Obviously, all remarks concerning the option
of choosing different kinetic terms for the lattice fermion theory apply equally to the
free massless quark case, and so there must be a whole surface in the space
of couplings for the generalized frustrated theory which all flow to the same fixed point.

As for real lattice QCD with quarks interacting with gluons,
it has been shown that at strong coupling the single-site colorless states are all degenerate
to leading order in $g^2$ and that this enormous degeneracy is lifted in order
$1/g^2$, see Ref.\cite{chiralsymmbb}.  Working to this order in the
doubled theory one obtains, exactly as in the case of the $1+!$-dimensional
theory, a generalized antiferromagnet.  Thus, we see that for this truncation algorithm
the free field theory and the strong coupling theory can be obtained from one another
by varying parameters in the general renormalized Hamiltonian.  This tells us that
the full generalized frustrated $SU(12)$ antiferromagnet theory has regions in which
the degrees of freedom are no longer free massless quarks, but are instead quarks interacting
with color gauge fields.  Note, adding the baryonic states which can be made
from $qqq$ configurations leads to a further generalization of the antiferromagnetic
system in which there is more than one irreducible representation of the
symmetry group for each site in the lattice and correspondingly more complicated
interactions.  All of these couplings become significant as one moves to
weak coupling but in the limit of very large $g^2$ the states with baryons in them
split away from the states with only mesons in order $1/g^2$.

Finally, returning to the point raised in the introduction; namely, although
this exercise was originally motivated by the desire to show that CORE provides
framework within which the old strong coupling treatment of QCD can be extended
to the weak coupling regime, at this point the question arises as to whether or
not one can tell if in fact quarks and gauge-fields are really fundamental.
More thought has to be given to the question of whether or not one can meaningfully
distinguish lattice QCD from the frustrated antiferromagnet.  Clearly much work
remains to be done to settle how various features of the various undoubled theories, such
as the anomaly, work in practice.  Also, it remains to be seen how all of these
undoubled theories differ from one another.

\end{document}